\documentclass{ws-mpla}

\begin{document}

\markboth{Hua-Xing Chen} {Scalar Tetraquark Currents With
Application to the QCD Sum Rule}

\catchline{}{}{}{}{}

\title{Scalar Tetraquark Currents with application to the QCD sum rule}

\author{\footnotesize Hua-Xing Chen}

\address{Research Center for Nuclear Physics, Osaka University, Ibaraki
567--0047, Japan
\\ Department of Physics, Peking University, Beijing
100871, China \\
hxchen@rcnp.osaka-u.ac.jp}

\author{Atsushi Hosaka}

\address{Research Center for Nuclear Physics, Osaka University, Ibaraki
567--0047, Japan}

\author{Shi-Lin Zhu}

\address{Department of Physics, Peking University, Beijing
100871, China}

\maketitle

\pub{Received (Day Month Year)}{Revised (Day Month Year)}

\begin{abstract}
We study the light scalar mesons in the QCD sum rule. We construct
both the diquark-antidiquark currents $(qq)(\bar q \bar q)$ and the
meson-meson currents $(\bar qq)(\bar qq)$. We find that there are
five independent currents for both cases, and derive the relations
between them. For the meson-meson currents, five independent
currents are formed by products of color singlet $\bar q q$ pairs or
color octet pairs. However, they can be related to each other, and
the relations are derived. We obtain the masses of the light scalar
mesons which are consistent with the experiments.

\end{abstract}

\ccode{PACS Nos.: 12.39.Mk, 12.38.Lg, 12.40.Yx.}

\section{Introduction}

Most of hadron states, including mesons and baryons, can be well
classified by the quark content $\bar q q$ and $qqq$ in the quark
model. However, there are still many observed states, which can not
be explained or difficult to be explained by just using $\bar q q$
and $qqq$. One important example is the pentaquark $\Theta^+$. After
three years of intense study, the status of $\Theta^+$ is still
controversial. The charm-strange mesons $D_{sJ}(2317)$,
$D_{sJ}(2460)$ and the charmonium state $X(3872)$, $Y(4260)$ are
also difficult to be explained by the conventional picture of $\bar
q q$ in the quark model.

Besides them, the light scalar mesons with masses below $1$ GeV have
been discussed for thirties year, but are still
controversial~\cite{Yao:2006px,Jaffe:1976ig}. By using the
conventional picture of $\bar q q$,  the mass ordering are expected
be $m_\sigma \sim m_{a_0} < m_\kappa < m_{f_0}$. However, from the
experiments, the mass ordering are $m_\sigma < m_\kappa < m_{f_0}
\sim m_{a_0}$~\cite{experiment}, which can be explained well
assuming that they were tetraquark states.

In this paper, we study the light scalar mesons in the QCD sum rule
employing tetraquark currents for the interpolating fields. We
construct both the diquark-antidiquark currents $(qq)(\bar q \bar
q)$ and the meson-meson currents $(\bar qq)(\bar qq)$. We find that
there are five independent currents for each scalar tetraquark
state, and derive the relations between them. We also find that
there are five meson-meson currents where ``mesons'' inside are
color singlets, and five ones where ``mesons'' inside are color
octets. However, they can be related to each other, and the
relations are obtained. The masses of the light scalar mesons are
calculated in the QCD sum rule, which are consistent with the
experiments.

\section{Diquark-Antidiquark Currents}

In this section, we construct the diquark-antidiquark currents
$(qq)(\bar q \bar q)$ for the state $\sigma(600)$. The currents for
other scalar mesons are similar. In order to make a scalar
tetraquark current, the diquark and antidiquark fields should have
the same color, spin and orbital symmetries. Therefore, they must
have the same flavor symmetry, which is either antisymmetric
($\mathbf{\bar 3_f} \otimes \mathbf{3_f}$) or symmetric
($\mathbf{6_f} \otimes \mathbf{\bar 6_f}$). In this paper we choose
the antisymmetric one. The details about their flavor structure can
be found in the reference~\cite{Chen:2007xr}.

Using the antisymmetric combination for diquark flavor structure, we
arrive at the following five independent currents
%
\begin{eqnarray}
\nonumber\label{define_udud_current} S^\sigma_3 &=& (u_a^T C
\gamma_5 d_b)(\bar{u}_a \gamma_5 C \bar{d}_b^T - \bar{u}_b \gamma_5
C \bar{d}_a^T)\, ,
\\ \nonumber
V^\sigma_3 &=& (u_a^T C \gamma_{\mu} \gamma_5 d_b)(\bar{u}_a
\gamma^{\mu}\gamma_5 C \bar{d}_b^T - \bar{u}_b \gamma^{\mu}\gamma_5
C \bar{d}_a^T)\, ,
\\
T^\sigma_6 &=& (u_a^T C \sigma_{\mu\nu} d_b)(\bar{u}_a
\sigma^{\mu\nu} C \bar{d}_b^T + \bar{u}_b \sigma^{\mu\nu} C
\bar{d}_a^T)\, ,
\\ \nonumber
A^\sigma_6 &=& (u_a^T C \gamma_{\mu} d_b)(\bar{u}_a \gamma^{\mu} C
\bar{d}_b^T + \bar{u}_b \gamma^{\mu} C \bar{d}_a^T)\, ,
\\ \nonumber
P^\sigma_3 &=& (u_a^T C d_b)(\bar{u}_a C \bar{d}_b^T - \bar{u}_b C
\bar{d}_a^T)\, ,
\end{eqnarray}
%
where the sum over repeated indices ($\mu$, $\nu, \cdots$ for Dirac,
and $a, b, \cdots$ for color indices) is taken. Either plus or minus
sign in the second parentheses ensures that the diquarks form the
antisymmetric combination in the flavor space. The currents $S$,
$V$, $T$, $A$ and $P$ are constructed by scalar, vector, tensor,
axial-vector, pseudoscalar diquark and antidiquark fields,
respectively. The subscripts $3$ and $6$ show that the diquarks
(antidiquark) are combined into the color representation
$\mathbf{\bar 3_c}$ and $\mathbf{6_c}$ ($\mathbf{3_c}$ or
$\mathbf{\bar 6_c}$), respectively.

\section{Meson-Meson Currents}

In this section, we construct the meson-meson currents $(\bar q
q)(\bar q q)$ for the state $\sigma(600)$. We find that there are
five currents where ``mesons'' inside are color singlets
%
\begin{eqnarray}\nonumber\label{define_meson_currents}
S^\sigma_1 &=& (\bar{u}_a u_a)(\bar{d}_b d_b) - (\bar{u}_a
d_a)(\bar{d}_b u_b)\, ,
\\ \nonumber
V^\sigma_1 &=& (\bar{u}_a\gamma_\mu u_a)(\bar{d}_b\gamma^\mu d_b) -
(\bar{u}_a\gamma_\mu d_a)(\bar{d}_b\gamma^\mu u_b) \, ,
\\
T^\sigma_1 &=&
(\bar{u}_a\sigma_{\mu\nu}u_a)(\bar{d}_b\sigma^{\mu\nu}d_b) -
(\bar{u}_a\sigma_{\mu\nu}d_a)(\bar{d}_b\sigma^{\mu\nu}u_b)\, ,
\\ \nonumber
A^\sigma_1 &=&
(\bar{u}_a\gamma_\mu\gamma_5u_a)(\bar{d}_b\gamma^\mu\gamma_5d_b ) -
(\bar{u}_a\gamma_\mu\gamma_5d_a)(\bar{d}_b\gamma^\mu\gamma_5u_b )\,
,
\\ \nonumber
P^\sigma_1 &=& (\bar{u}_a\gamma_5u_a)(\bar{d}_b\gamma_5d_b) -
(\bar{u}_a\gamma_5d_a)(\bar{d}_b\gamma_5u_b)\, .
\end{eqnarray}
%
The minus sign ensures that the diquarks (anti-diquarks) form the
antisymmetric combination in the flavor space. These five currents
are independent, and can be related to the five diquark-antidiquark
currents
%
\begin{eqnarray}\nonumber\label{relations1}
8 S^\sigma_3 &=& - 2 S^\sigma_1 - 2 V^\sigma_1 + T^\sigma_1 - 2
A^\sigma_1 - 2 P^\sigma_1 \, ,
\\ \nonumber
2 V^\sigma_3 &=& 2 S^\sigma_1 - V^\sigma_1 + A^\sigma_1 - 2
P^\sigma_1 \, ,
\\
2 T^\sigma_6 &=& 6 S^\sigma_1 + T^\sigma_1 + 6 P^\sigma_1 \, ,
\\ \nonumber
2 A^\sigma_6 &=& 2 S^\sigma_1 + V^\sigma_1 - A^\sigma_1 - 2
P^\sigma_1\, ,
\\ \nonumber
8 P^\sigma_3 &=& - 2 S^\sigma_1 + 2 V^\sigma_1 + T^\sigma_1 + 2
A^\sigma_1 - 2 P^\sigma_1 \, .
\end{eqnarray}
%
We find the other five currents where ``mesons'' inside are color
octets
%
\begin{eqnarray}\nonumber\label{define_meson_currents}
S_8 &=& (\bar{u}_a{\lambda^n_{ab}}u_b)(\bar{d}_c{\lambda^n_{cd}}d_d)
- (\bar{u}_a{\lambda^n_{ab}}d_b)(\bar{d}_c{\lambda^n_{cd}}u_d)\, ,
\\ \nonumber
V_8 &=& (\bar{u}_a\gamma_\mu {\lambda^n_{ab}}
u_b)(\bar{d}_c\gamma^\mu {\lambda^n_{cd}} d_d) -
(\bar{u}_a\gamma_\mu {\lambda^n_{ab}} d_b)(\bar{d}_c\gamma^\mu
{\lambda^n_{cd}} u_d)\, ,
\\ T_8 &=& (\bar{u}_a\sigma_{\mu\nu}
{\lambda^n_{ab}} u_b)(\bar{d}_c\sigma^{\mu\nu} {\lambda^n_{cd}} d_d)
- (\bar{u}_a\sigma_{\mu\nu} {\lambda^n_{ab}}
d_b)(\bar{d}_c\sigma^{\mu\nu} {\lambda^n_{cd}} u_d)\, ,
\\ \nonumber A_8 &=& (\bar{u}_a\gamma_\mu\gamma_5 {\lambda^n_{ab}}
u_b)(\bar{d}_c\gamma^\mu\gamma_5 {\lambda^n_{cd}} d_d) -
(\bar{u}_a\gamma_\mu\gamma_5 {\lambda^n_{ab}}
d_b)(\bar{d}_c\gamma^\mu\gamma_5 {\lambda^n_{cd}} u_d)\, ,
\\ \nonumber P_8 &=& (\bar{u}_a\gamma_5 {\lambda^n_{ab}}
u_b)(\bar{d}_c\gamma_5 {\lambda^n_{cd}} d_d) - (\bar{u}_a\gamma_5
{\lambda^n_{ab}} d_b)(\bar{d}_c\gamma_5 {\lambda^n_{cd}} u_d)\, .
\end{eqnarray}
%
They are also independent, and can be related to the five
diquark-antidiquark currents, as well as to the five meson-meson
currents $S^\sigma_1$, $V^\sigma_1$, $T^\sigma_1$, $A^\sigma_1$ and
$P^\sigma_1$
%
\begin{eqnarray}\nonumber\label{relations1}
12 S^\sigma_8 &=& - 2 S^\sigma_1 + 6 V^\sigma_1 + 3 T^\sigma_1 - 6
A^\sigma_1 - 6 P^\sigma_1 \, ,
\\ \nonumber
3 V^\sigma_3 &=& 6 S^\sigma_1 - 5 V^\sigma_1 - 3 A^\sigma_1 - 6
P^\sigma_1 \, ,
\\
3 T^\sigma_6 &=& 18 S^\sigma_1 - 5 T^\sigma_1 + 18 P^\sigma_1 \, ,
\\ \nonumber
3 A^\sigma_6 &=& - 6 S^\sigma_1 - 3 V^\sigma_1 - 5 A^\sigma_1 + 6
P^\sigma_1\, ,
\\ \nonumber
12 P^\sigma_3 &=& 6 S^\sigma_1 - 6 V^\sigma_1 + 3 T^\sigma_1 + 6
A^\sigma_1 - 2 P^\sigma_1 \, .
\end{eqnarray}
%

%
\section{QCD sum rule analysis}\label{sec_sumrule}
%

We have performed the QCD sum rule analysis for each single current
and their linear combinations. We have performed the OPE calculation
up to dimension eight, which contains the four-quark condensates. We
find that the results for single currents are not always reliable,
while a good sum rule is achieved by a linear combination of
$A_6^\sigma$ and $V_3^\sigma$
%
\begin{eqnarray}\label{eq_eta_sigma_1}
\eta^\sigma &=& \cos\theta A^\sigma_6 + \sin\theta V^\sigma_3 \, ,
\end{eqnarray}
%
where $\theta$ is the mixing angle. The best choice of the mixing
angle turns out to be $\cot\theta = 1/\sqrt{2}$. The mixed currents
for $\kappa$, $a_0$ and $f_0$ can be found in the similar way.

By using this mixed current $\eta^\sigma$, we studied Borel mass
$M_B$ and threshold value $s_0$ dependences, which are quite stable.
The convergence of the OPE is also good with the positivity of the
spectral densities being maintained, and with sufficient pole
contribution. Therefore, we have achieved a good QCD sum rule within
the present calculation of OPE. We also considered the finite decay
width by using the Gaussian distribution instead of the pole term in
the phenomenological side, where the predicted masses do not change
much as far as the Borel mass is within a reasonable range. Then we
can still reproduce the experimental data.

We have also performed the QCD sum rule analysis with the
conventional $\bar q q$ currents. Their masses are calculated to be
around 1.2 GeV as in the previous work~\cite{Reinders:1981ww}. This
indicates that the tetraquark currents are more suitable for the
description of the light scalar mesons than the conventional ones.

In summary, our QCD sum rule analysis supports a tetraquark
structure for low-lying scalar mesons. We construct both the
diquark-antidiquark currents and the meson-meson currents. We find
that there are five independent currents in both constructions.
However, currents in different constructions can be related to each
other. Therefore, all the scalar tetraquark currents can be written
as a combination of five meson-meson currents where ``mesons''
inside are color singlets. This conclusion can be extended to other
tetraquark currents of different quantum numbers, as well as
pentaquark currents.

\section*{Acknowledgments}

H.X.C. is grateful for Monkasho support for his stay at the Research
Center for Nuclear Physics where this work is done. A.H. is
supported in part by the Grant for Scientific Research ((C)
No.19540297) from the Ministry of Education, Culture, Science and
Technology, Japan. S.L.Z. was supported by the National Natural
Science Foundation of China under Grants 10625521 and 10721063 and
Ministry of Education of China.

\end{document}